\documentclass[a4paper,12pt]{article}

\usepackage[english]{babel}
\usepackage[utf8x]{inputenc}
\usepackage[T1]{fontenc}
\usepackage[flushmargin]{footmisc}
\usepackage{setspace}
\usepackage[comma]{natbib}
\usepackage{float}
\usepackage{amsmath}
\usepackage{amsfonts}
\usepackage{amssymb}
\usepackage{ae}
\usepackage{tabularx}
\usepackage{caption}
\usepackage{subfigure}

\usepackage[a4paper,top=3cm,bottom=2cm,left=3cm,right=3cm,marginparwidth=1.75cm]{geometry}

\usepackage{graphicx}
\usepackage[colorinlistoftodos]{todonotes}
\usepackage[colorlinks=true, allcolors=blue]{hyperref}
\def\ci{\perp\!\!\!\perp}

\title{What Is the Value Added by Using Causal Machine Learning Methods in a Welfare Experiment Evaluation? }

\author{Anthony Strittmatter \\ CREST-ENSAE } %\\  First Draft: September 17, 2018}

\begin{document} \maketitle   \thispagestyle{empty}

\begin{abstract}
Recent studies have proposed causal machine learning (CML) methods to estimate conditional average treatment effects (CATEs). In this study, I investigate whether CML methods add value compared to conventional CATE estimators by re-evaluating Connecticut's Jobs First welfare experiment. This experiment entails a mix of positive and negative work incentives. Previous studies show that it is hard to tackle the effect heterogeneity of Jobs First by means of CATEs. I report evidence that CML methods can provide support for the theoretical labor supply predictions. Furthermore, I document reasons why some conventional CATE estimators fail and discuss the limitations of CML methods.
\end{abstract}


\vfill

\noindent \rule{8cm}{0,2mm} \\
\noindent {\footnotesize The Manpower Demonstration Research Corporation (MDRC) provided the experimental data used in this article. I remain solely responsible for how the data have been used and interpreted. The study is part of the Swiss National Science Foundation (SNSF) project ``Causal Analysis with Big Data'', grant number SNSF 407540\_166999, and is included in the Swiss National Research Program ``Big Data'' (NRP 75). A previous version of the paper was presented at Erasmus University Rotterdam, UC Berkeley, University of Amsterdam, University of Munich. I thank participants, particularly Ulrich Glogowsky, Bryan Graham, Pat Kline, Michael Knaus, and Michael Lechner, for helpful comments and suggestions. The usual disclaimer applies.  
Address for correspondence: Anthony Strittmatter, CREST-ENSAE, 5 Av Henry Le Chatelier, 91120 Palaiseau, France, \linebreak \href{mailto:Anthony.Strittmatter@ensae.fr}{Anthony.Strittmatter@ensae.fr},  \url{www.anthonystrittmatter.com}.

\clearpage \setcounter{page}{1}
 \pagestyle{plain}
\onehalfspacing \normalsize

\section{Introduction}

Causal machine learning (CML) methods are new in the toolbox of economists \citep[see, e.g.,][for a review]{ath19b}.
They are promising for the estimation of heterogeneous policy effects conditional on exogenous covariates, which are often labelled conditional average treatment effects (CATEs). Compared to more conventional estimation methods, CML methods have three potential advantages. First, they make it convenient to incorporate many covariates that are potentially responsible for effect heterogeneity. Several conventional estimation approaches also make it possible to incorporate many covariates, but CML methods can avoid in a data-driven way the potential risk of overfitting and are computationally feasible even when the covariate space is very large. %Some CML methods allow for more covariates than observations. 
Second, CML methods are relatively flexible when dealing with the covariates. Some CML methods incorporate nonlinear and interaction terms automatically. Third, the systematic CML algorithms make it less likely to overlook important effect heterogeneity. However, judiciously conducted conventional approaches may also find important heterogeneity margins. Furthermore, CML methods are largely black-box approaches, which is certainly a disadvantage. Accordingly, it is unclear how much value CML methods can add to economic applications compared to more conventional estimation methods.

In this study, I revisit the effects of Connecticut's Jobs First welfare experiment on the labor supply. Well-established labor supply theory provides clear predictions about the heterogeneity margins of this experiment \citep[see, e.g.,][for a comprehensive summary]{kli16a}. However, \cite{bit17} document the limitations of a conventional CATE estimator in terms of its ability to provide evidence for these theoretical predictions. This is puzzling because \cite{bit06} show that quantile treatment effects (QTEs) can uncover evidence for the theoretical labor supply model. It appears that the Jobs First data contain relevant information that can support labor supply theory, but how to uncover the appropriate heterogeneity by means of conventional CATE estimators is not straightforward.   

This study contributes to the aforementioned literature in at least three ways. First, I investigate whether CATEs estimated with CML methods can provide evidence supporting the theoretical labor supply predictions of the Jobs First program, which would clearly represent value added compared to conventional CATE estimators. Second, I reveal modeling restrictions that prevent the conventional CATE estimators from revealing more effect heterogeneity. Third, I test whether the estimates of the CATEs and QTEs are nested.

\cite{bit17} consider local constant models, which are one of the workhorse methods used to estimate CATEs in empirical economics. Local constant models stratify the sample into different groups defined by the covariates and report subgroups' average treatment effects. Local constant models uncover effect heterogeneity across groups but report constant effects within groups. There are three potential reasons why local constant models fail to support labor supply theory. First, the choice of the subgroups could be suboptimal. Second, constant effects within groups do not accurately approximate the possibly continuously distributed treatment effects. Third, the covariates used in the local constant models may be insufficient for explaining effect heterogeneity.\footnote{Obviously, local constant models would also fail if the theoretical predictions are not good approximations of the labor supply effects or if measurement error or other data problems prevent me from finding empirical support for the labor supply predictions.} Suitable CML methods can overcome all three possible disadvantages of local constant models. It is essential to understand why local constant models fail to support theoretical predictions in the Jobs First case because these models are widely used, hence, the disadvantages may carry over to other applications. 

For the main analysis, I use the double machine learning approach proposed in \cite{che17}. This is a generic approach that can incorporate many different machine learning estimators. Accordingly, it is suitable to compare machine learning estimators with different modeling restrictions.
% I consider machine learning estimators that can be used with standard personal computers. 
%These machine learning estimators can be accessed by a widespread audience. Furthermore, these e no special IT infrastructure is required, in contrast to machine learning methods that require cloud computing infrastructure. 
%This is useful when dealing with confidential data, such as data from the Manpower Demonstration Research Corporation (MDRC), because it is often easier to comply with data security regulations and laws using a personal computer than on distributed computing systems.
I consider the ``tree'' and the ``random forest'' machine learning estimators \citep[see, e.g.,][]{hast09}. Tree estimators split the data into mutually exclusive groups defined by the covariates and report effect heterogeneity as the subgroups' average treatment effects. Similar to local constant models, tree estimators uncover effect heterogeneity across groups but report constant effects within groups. However, tree estimators employ data-driven algorithms to select subgroups, whereas for local constant models, subgroups must be manually selected. Both methods could, in principle, incorporate many covariates. However, it is more convenient to use tree estimators when the covariate space is large. In particular, trees can automatically incorporate different subgroup definitions (based on covariates and interactions between covariates) without precoding. Random forest estimators are ensemble methods. They estimate many trees based on different subsets of the data and covariates and then report the average of the different tree estimates. These features imply that random forest estimators additionally relax the restriction of constant effects within subgroups. Accordingly, the tree and random forest estimators are suitable for relaxing, in a stepwise fashion, the modeling restrictions of local constant models. 

The results suggest that random forest estimators can provide evidence for the theoretical labor supply model when they incorporate many covariates. This suggests that including many relevant covariates and allowing for continuously distributed treatment effects are important ingredients for establishing a match between theoretical predictions and empirical results in the Jobs First application. Using solely data-driven methods of selecting subgroups, without incrementing the covariates and flexibility of the model, does not seem to improve the matches. 

Furthermore, the results provide evidence that the CML estimates of the CATEs and QTEs provide disparate sets of information for evaluating the Jobs First case, suggesting that the estimated CATEs do not uncover all of the inherent effect heterogeneity of the Jobs First experiment. Both CATEs and QTEs have advantages and disadvantages. CATEs could be useful to design welfare schemes that optimize the labor supply response of specific target groups or to create assignment rules. QTEs enable the study of the responses of the entire labor supply distribution (in a fully flexible way), but it is difficult to assign these responses to specific groups.

Using case studies to demonstrate the capabilities of new methods is common in economics. 
Recently, \cite{kl18} illustrate how predictive machine learning methods can improve human decisions, using the example of bail decisions made by judges. \cite{do18} launched a causal inference data analysis challenge. Contributors received real-world data to estimate the effects of birth weight on child's IQ. The real-world data were slightly calibrated such that the ground truth was known by the organizers of the challenge. Their main conclusion is that flexible methods with fewer modeling restrictions perform better, which is coherent with my findings for the Jobs First application.

The number of economic applications using CML methods to predict CATEs is rapidly increasing. \cite{dav17} estimate the heterogeneous effects of summer jobs on the probability of committing a violent crime. \cite{tadd16} investigate the heterogeneous effects of A/B experiments in online-auctions (eBay) on customer responses. \cite{ber17} estimate the heterogeneous effects of a work experience program in C\^{o}te d'Ivoire on post-participation employment and wages. \cite{kna18b} estimate the heterogeneous
employment effects of a job search program in Switzerland. \cite{farb21} estimate heterogeneous impacts of financial circumstance on cognitive performance. \cite{and18}, \cite{cag21}, \cite{haus21}, and \cite{kni19} develop CATE-based targeting rules for economic policies. 

In the next section, I provide some background information about the Jobs First welfare experiment. In Section \ref{sec3}, I introduce the MDRC data. In Section \ref{sec4}, I describe the empirical framework of this study. In Section \ref{sec5}, I document the empirical results. The final section concludes. The Online Appendices A-F provide supplementary material and descriptive statistics.

\section{The Jobs First welfare program \label{sec2}}

In 1996, Connecticut replaced the Aid for Families with Dependent Children (AFDC) program with the Jobs First program. The Jobs First program created financial work incentives for people on assistance that differed from those offered by the AFDC program. 

Figure \ref{fig0} shows the earnings and welfare transfers in a stylized way. The maximum welfare payment W is similar under both welfare schemes, but additional earnings lead to different welfare payment deductions. AFDC disregarded all earnings below a fixed amount B, which was \textdollar 120 per month during the first 12 months of employment while on assistance and \textdollar 90 per month afterward. Furthermore, 51\% of any additional earnings was disregarded during the first four months of employment and 27\% of any additional earnings afterward. By contrast, the Jobs First program disregards all earnings below the federal poverty line (FPL). Earnings above the FPL terminate all welfare benefit payments from the Jobs First program (which is like a cliff in the benefits payment scheme). The Jobs First and AFDC programs differ in other aspects besides the financial work incentives. The additional changes are summarized in the Online Appendix A.

\[ \text{Figure~\ref{fig0} around here}  \]

\cite{bit06} use a static labor supply model to develop four hypotheses. First, the Jobs First program deducts fewer earnings from welfare payments than the AFDC program. Thus, Jobs First should have a positive effect on the extensive margin of the labor supply. Second, Jobs First recipients with relatively low earnings (between B and E in Figure \ref{fig0}) can keep more of their additional income than AFDC recipients. This should create positive work incentives when the substitution effects dominate the income effects. Third, the FPL is considerably higher than the earnings amount E at which participants lose their eligibility for AFDC welfare benefit payments. Accordingly, this provides a lump-sum transfer to Jobs First welfare recipients with earnings between E and the FPL, which reduces the optimal earnings in the presence of negative income effects. Furthermore, the cliff construction of the Jobs First payment scheme creates incentives to reduce earnings to just below the FPL, which might increase income or leisure time. Fourth, for recipients sufficiently above the FPL, the AFDC and Jobs First programs provide the same labor supply incentives. 

The theoretical predictions and an additional data restriction lead to three empirically testable hypotheses: 
\begin{itemize}
\item[H1:] The Jobs First program has positive earnings effects for some individuals with zero earnings under AFDC. Furthermore, the earnings effects cannot be negative for this group because the earnings outcome cannot be negative (additional data restriction).
\item[H2:] There is a mix of positive and negative earnings effects in the group of individuals with positive earnings below the FPL under AFDC.
\item[H3:] Jobs First has non-positive earnings effects for individuals with earnings above the FPL under AFDC.
\end{itemize}

\section{Experimental data \label{sec3}}

The Connecticut Department of Social Services required the  Manpower Demonstration Research Corporation (MDRC) to conduct a randomized controlled trial to evaluate the Jobs First program. Experimental participants were single-parent welfare applicants and recipients who lived in Manchester or New Haven. Between January 1996 and February 1997, 4,803 experimental participants were randomly assigned to either the AFDC (control group) or Jobs First (treatment group) programs.\footnote{I drop one experimental participant who had extraordinarily high earnings. This does not change the point estimates much but makes the estimation of the confidence intervals more stable.}  

\subsection{Variable definitions}

The MDRC's public use files for the Jobs First program contain baseline data on demographic
and family composition variables merged with longitudinal administrative information
on welfare and food stamps payments and earnings provided by the state unemployment insurance
system. The outcome variable is earnings per quarter in US dollars.\footnote{I do not observe the earnings reported by the experimental participants to the welfare agency, which could matter when misreporting is a major practice \citep[see discussion, e.g., in][]{kli16a}.} The treatment is assignment to the Jobs First program. I follow 4,802 experimental participants for seven quarters after random assignment (RA) to the Jobs First program or AFDC program. The total sample contains 33,614 observations.

Being above or below the FPL is one of the major factors driving effect heterogeneity according to the theoretical considerations. Similar to previous Jobs First studies, I cannot observe the administrative assistance unit size, which determines the FPL of the experimental participant. Following the suggestions of \cite{kli16a}, I calculate the assistance unit size based on the number of children at RA.\footnote{I cannot calculate the assistance unit size for 160 experimental participants because the number of children is not reported.} This approach may lead to the underestimation of the FPL because mothers may have more children during the seven quarters after RA. To account for this potential disadvantage, I inflate the number of children by one for all mothers as a robustness check (in the following, I call this calculation of the FPL ``extra child'').

I distinguish between three sets of covariates that I use for the heterogeneity analysis. I label them ``baseline'', ``decent'', and ``kitchen sink'' covariates. Table \ref{list} summarizes the different covariate categories. The baseline covariates contain the elapsed quarters since RA and the earnings in the seven quarters prior to RA. The selection of the baseline covariates follows \cite{bit17}, who use these two covariates in some of their main specifications. 

\[ \text{Table~\ref{list} around here}  \]

The decent covariates include 13 variables. In addition to the baseline variables, they include age, education, information about children, and more information about earnings and welfare history. This set of covariates is still relatively small; however, it may be difficult to consider even these relatively few variables with a standard CATE estimator when including many non-linearities and interactions between the covariates. 

The kitchen sink covariates include 68 variables. I include all exogenous covariates for which data are available from the MDRC and that might affect effect heterogeneity. The kitchen sink covariates include different measures of the variables that are already included in the decent covariates. The additional variables are ethnicity, marital status, information about the residence, information on previous participation in education or labor market programs, and more information on earnings and welfare history.

\subsection{Descriptive statistics}
Table \ref{desc} reports the descriptive statistics of the main variables. The mean earnings do not differ greatly between the Jobs First and AFDC participants. Thus, the average effects of the Jobs First experiment do not differ significantly from zero \citep[similar to previous findings, e.g., in][]{bit06}. However, fewer participants in the Jobs First program are unemployed compared to the participants in the AFDC program. More participants in the Jobs First program have employment with earnings below the FPL than participants in the AFDC program. These descriptive findings are in line with the theoretical labor supply predictions. The group of participants with earnings above the FPL is relatively small regardless of whether I inflate the number of children used to calculate the FPL or not. The share of participants with earnings above the FPL does not differ much between the Jobs First and AFDC participants.

Table \ref{desc} documents the standardized difference in the baseline and decent covariates between the Jobs First and AFDC participants.\footnote{The standardized difference in variable $X$ between samples $A$ and $B$ is defined as 
\begin{equation*}
SD = \frac{|\bar{X}_A-\bar{X}_B|}{\sqrt{\frac{1}{2}\left(Var(X_A) + Var(X_B)\right)}}\cdot 100,
\end{equation*}
where $\bar{X}_A$ denotes the mean of sample $A$ and $\bar{X}_B$ denotes the mean of sample $B$. \cite{ro83} consider an absolute standardized difference higher than 20 to be ``large.''} Table B.1 in the Online Appendix B shows the descriptive statistics of the kitchen sink covariates. If the RA to the Jobs First and AFDC programs was appropriately random, then we expect all pre-RA covariates to be balanced. Table \ref{desc} shows that there are no large differences between the pre-RA covariates. However, there are some small differences. Jobs First participants have slightly more children, less previous earnings, and received more welfare than AFDC recipients.

\[ \text{Table~\ref{desc} around here}  \]

\section{Empirical approach \label{sec4}}
\subsection{Estimation target}

The treatment dummy $D_i$ equals one when an experimental participant is assigned to Jobs First and zero when she is assigned to the AFDC welfare scheme. Following Rubin's (1974) \nocite{rub74} potential outcome framework, $Y_{it}(1)$ denotes the potential earnings outcome under Jobs First for individual $i$ in quarter $t$ (for $i=1,…,N$ and $t = 1, ..., 7$). Correspondingly, $Y_{it}(0)$ denotes the potential earnings outcome under AFDC for individual $i$ in quarter $t$. Each individual can be assigned to either the Jobs First or AFDC but not to both welfare schemes simultaneously. Thus, only one potential outcome is observable. Under the stable unit treatment value assumption (SUTVA), the observed outcome equals
\begin{equation} \label{eq1}
Y_{it} = Y_{it}(1) D_i + Y_{it}(0)(1-D_i).
\end{equation}

Individual $i$'s causal effect of being assigned to Jobs First instead of the AFDC welfare scheme on earnings is
\begin{equation*}
\delta_{it} = Y_{it}(1) -Y_{it}(0).
\end{equation*}
We cannot identify individual causal effects without assumptions that are implausible in many applications (e.g., the assumption of effect homogeneity). Nevertheless, group averages of $\delta_{it}$ may be identifiable under plausible assumptions. For example, the identification of the average treatment effect (ATE), $\rho=E[\delta_{it} ]$, and the average treatment effect on the treated (ATET), $\theta=E[\delta_{it} |D_i=1]$, is standard in policy evaluations \citep[see, e.g.,][]{imb09}. CATEs can potentially uncover effect heterogeneity based on exogenous pre-treatment variables $X_{it}$. 
The CATEs are
\begin{equation*}
\bar{\delta}(x) = E[\delta_{it}|X_{it}=x]. 
\end{equation*}

Under the random treatment assignment and SUTVA, the CATEs
\begin{equation*}
\bar{\delta}(x) =  E[Y_{it} |D_i=1,X_{it}=x]-E[Y_{it}|D_i=0,X_{it}=x],
\end{equation*}
are identified from observable data on $(Y_{it},D_i,X_{it})$.

The CATEs are often labeled as individualized or personalized treatment effects. To some extent, this is misleading because these labels might suggest that the CATEs closely approach the individual causal effects. However, to achieve this, the individual causal effects must be (almost) deterministic, and all relevant determining variables must be observed. In many applications, these requirements are too strong. Nevertheless, even when the CATEs are not equal to the individual causal effects, they have the potential to provide a more complete picture of the effect heterogeneity than the ATEs can.

\subsection{Local constant model }

The local constant model partitions the sample into mutually exclusive groups. Let $\pi= \{ g_1, ...,g_{\# (\pi)} \}$ be a specific sample partition, let $g_j \equiv g_j (x,\pi)$ be the respective group (for $j=1,…,\#(\pi))$, and let $\#(\pi)$ be the number of groups in the partition $\pi$. The group $g_j (x,\pi)$ of partition $\pi$ is a function of the covariate space of $X_{it}$. % such that $x \in g_j(x,\pi)$. 
For an explicit example, consider that $X_{it}$ contains only a binary indicator for gender. Then, we can choose between two possible sample partitions; either keep men and women together, $\pi'=\{ g_1 \}=\{ \mbox{men}, \mbox{women} \}$, or we partition men and women into two separate groups, $\pi'' =\{ g_1,g_2 \}= \{ \{ \mbox{men} \}, \{ \mbox{women} \} \}$.  

Local linear models can be estimated with a linear model that incorporates the interaction terms with $D_i$
\begin{equation} \label{lcm}
Y_{it} =  \sum_{j=1}^{\#(\pi)} \gamma_{j} \cdot 1\{ X_{it} \in g_j(x,\pi)\}+  \sum_{j=1}^{\#(\pi)} \delta_{j} \cdot D_i \cdot 1\{ X_{it} \in g_j(x,\pi)\}.
\end{equation}
The CATEs are $\bar{\delta}(X_{it})= \sum_{j=1}^{\#(\pi)} \delta_{j} \cdot 1\{ X_{it} \in g_j(x,\pi)\}$.

The empirical challenge is to find a useful sample partition $\pi$. Often the choice of $\pi$ is a rather discretionary decision. Not accounting for the selection of $\pi$ (which may include many manual model selection steps) can lead to invalid inference procedures.

\subsection{CML approach}

Given the discussion in the last section, it would be useful to select $\pi$ in a data-driven way based on some optimality criteria. In the ideal case, these optimality criteria would be based on the individual causal effect $\delta_{it}$. However, the fundamental problem of causal analysis is the unobservability of $\delta_{it}$. 

To overcome the fundamental identification problem, a popular CML approach is to modify the outcome. For example, I could replace the outcome with the orthogonal score
\begin{equation*}
Y_{it}^{*} = \mu_1(X_{it}) - \mu_0(X_{it}) + \frac{D_{i}(Y_{it} - \mu_1(X_{it}) )}{p(X_{it})} - \frac{(1-D_i)(Y_{it} - \mu_0(X_{it}) )}{1-p(X_{it})},
\end{equation*}
which goes back to \cite{robi95}. It includes the three nuisance parameters $\mu_1(X_{it}) = E[Y_{it}|D_i=1,X_{it}]$, $\mu_0(X_{it}) = E[Y_{it}(0)|X_{it}]=E[Y_{it}|D_i=0,X_{it}]$, and $p(X_{it})=E[D_{i}|X_{it}]$ that model the potential outcomes and the selection into treatment. Each of the nuisance parameters can be estimated with methods suited to making predictions, such as machine learning estimators. The expected value of the efficient score is the ATE, $\rho = E[Y_{it}^{*}]$, and the conditional expectations are the CATEs, $\bar{\delta}(x) = 
E[Y_{it}^{*}|X_{it} = x]$ (see the proof in the Online Appendix C). The orthogonal score has the advantage that the causal effect estimates remain consistent even when either $\mu_0(x)$ and $\mu_1(x)$ or $p(x)$ is misspecified.

Chernozhukov et al. (2018)\nocite{che17} call this approach the double machine learning approach because
it combines first-step auxiliary predictions of nuisance parameters to estimate the causal effects in the second step. These authors discuss how it is possible to obtain $\sqrt{N}$-consistent and asymptotically normal estimates of the ATE and other low-dimensional causal parameters. An important finding is that, even if the estimates of the nuisance parameter have a slow convergence rate (e.g., $\sqrt[4]{N}$), the ATE estimates can still converge with $\sqrt{N}$. However, much less is known about the asymptotic properties of the modified outcome approach regarding the functions of the efficient score, such as the CATEs \citep[see discussion, e.g., in][]{chern18b}. \cite{Lee2017} discuss the asymptotic properties of CATEs estimated with the orthogonal score, but they assume that the covariates used to model effect heterogeneity are low-dimensional and that the nuisance parameters are estimated at a parametric rate.

Another popular CML approach is the causal forest estimator \citep[see][]{Athey2017a,Wager2017}. Causal forest estimators are consistent and asymptotically normal for the estimation of CATEs, but the convergence rates are below $\sqrt{N}$. The modified outcome and causal forest approaches have similar good finite sample properties for estimating CATEs \citep[see, e.g.,][and references therein]{kna18}. I use the modified outcome approach in the main specifications. Additionally, I show that the results do not change considerably when using causal forest estimators.

Chernozhukov et al. (2018) suggest using a cross-fitting procedure to break through the correlation structure between the estimated nuisance parameters and the causal effect estimation. To implement this procedure, I partition the data into two random samples. I use the first sample to estimate the nuisance parameters and extrapolate the fitted values of the nuisance parameters to the second sample. Then, I use the second sample to estimate the CATEs.\footnote{For the forest estimators, I additionally switch the first and second samples and repeat the cross-fitting procedure. Then, I report the average CATEs obtained from the first and second samples. In the clustered bootstrap procedure that I use to compute the p-values and confidence intervals, I ensure that each individual can enter only one cross-fitting sample and never both.}

\subsection{Machine learning estimators}
Many different machine learning estimators can be combined with the modified outcome approach \citep[see, e.g.,][for an overview of different machine learning estimators]{hast09}. I focus on the regression tree and random forest estimators \citep[e.g.,][]{brei01} because they mimic the modeling restrictions of the local constant model. I use the R packages \texttt{rpart} and \texttt{grf} to implement those estimators.

\subsubsection{Tree estimator}

Similar to the local constant model, regression trees partition the sample into mutually exclusive groups $g_j$, which are now called leaves. For a specific sample partition $\pi$, which is now called a tree, I can estimate the CATEs by
\begin{equation*}
\hat{\bar{\delta}}(x,\pi) = \frac{1}{\sum_{i=1}^{N}\sum_{t=1}^{7} 1\{ X_{it} \in g_j(x,\pi)\}}\sum_{i=1}^{N}\sum_{t=1}^{7} 1\{ X_{it} \in g_j(x,\pi)\} \cdot \hat{Y}_{it}^{*},
\end{equation*}
where $\hat{Y}_{it}^{*}$ is the estimate of $Y_{it}^{*}$, which I extrapolate from the retained cross-fitting sample. 

Regression trees select the partition $\pi$ with a greedy algorithm, i.e., by adding recursive sample splits to the tree without anticipating later splits \citep[e.g.,][]{brei84}. Using the modified outcome approach, regression trees seek to minimize the mean squared error (MSE) with regard to $\hat{Y}_{it}^{*}$. Accordingly, they select the splits that fit the approximated CATEs best since the group average of $\hat{Y}_{it}^{*}$ approximates the CATEs. The first splits contribute more effect heterogeneity than the last splits, because of the hierarchical partition structure. I select the optimal tree $\pi^*$ and $\bar{\delta}_{\mbox{tree}}(x) = \delta(x,\pi^*)$ based on the out-of-sample MSE, which I calculate with a 10-fold cross-validation procedure. To stabilize the trees, I impose the restriction that each leaf should contain at least 50 observations.

Following the suggestions of \cite{ath16}, I use the so-called honest inference procedure, which means that I split the sample into two parts of equal size. Then, I use the first partition to build the tree (training sample) and the second partition to estimate the CATEs (estimation sample).\footnote{In the clustered bootstrap procedure that I use to compute p-values and confidence intervals, I ensure that each individual can enter either the training or estimation sample, but never both.} This separation between the training and estimation samples avoids overfitting of the estimated CATEs.

\subsubsection{Generalized random forest estimator}

Generalized random forests are assembled from $H$ decorrelated honest trees $\delta(x,\pi_h)$ (for $h = 1, ..., H$). The decorrelated honest trees are estimated using different subsamples of the data and subsets of the covariates. Decorrelation is necessary because, without it, each tree would have a similar structure and I would not be able to gain much from assembling the trees. The random forest estimator of the CATEs is the average of these honest trees: 
\begin{equation*}
\hat{\bar{\delta}}_{RF}(x) = \frac{1}{H} \sum_{h=1}^{H} \hat{\bar{\delta}}_{\mbox{tree}}(x,\pi_h).
\end{equation*}
The honest trees of a random forest are built deep (i.e., with small terminal leaves). Thus, I no longer try to optimize the leaf size of the trees with the cross-validation procedure. Instead, I build many deep honest trees that have a small bias but a large variance. Averaging across different honest trees reduces the variance (which is often called ``bagging''). \cite{Athey2017a} explore the consistency and asymptotic normality of generalized random forests.

I build random forests with 1,000 decorrelated trees, each with a minimum leaf size of 10 observations.\footnote{The number of trees $H$ is an important tuning parameter for random forests. Table E.1 in the Online Appendix E shows that the out-of-sample MSE improves when I increase the number of trees. However, with 100 trees, the prediction power of random forests is already almost saturated, and the MSE improvements are marginal.} In each subsample, I randomly select 50\% of the individuals and two-thirds of the covariates. 

\subsection{Testing the theoretical hypotheses}

To provide evidence for the theoretical predictions, I want to test whether the estimated CATEs are non-positive or non-negative in specific subsamples. From the forest estimators, I obtain separate CATE estimates for each individual. Imposing single hypothesis tests on each CATE would cause the multiple hypothesis testing problem. %Standard F-tests would have, in the worst case, no degrees of freedom. 
To avoid this problems, I employ discrete versions of first-order stochastic dominance tests and compute p-values using a clustered bootstrap procedure \citep[e.g.,][]{an96,bar03}. First, I test the null hypothesis $H_0^{+}$ that all estimated CATEs are non-negative. Second, I test the null hypothesis $H_0^{-}$ that all estimated CATEs are non-positive. I provide the details of the distributional tests in the Online Appendix D.

For the tests, I restrict the sample to the AFDC participants, which does not alter the expected CATE estimates because of the random treatment assignment. However, it enables the identification of CATEs at specific earnings levels under AFDC. According to the theoretical labor supply predictions, the estimated CATEs should
\begin{enumerate}
\item[H1:] reject $H_0^-$ and not reject $H_0^+$ in the subsample of unemployed AFDC participants,
\item[H2:] reject $H_0^+$ and $H_0^-$ in the subsample of AFDC participants with positive earnings below FPL, and
\item[H3:] reject $H_0^+$ and not reject $H_0^-$ in the subsample of AFDC participants with earnings above the FPL.
\end{enumerate}
These are necessary but not sufficient conditions of the labor supply predictions.

\subsection{Comparison between CATEs and QTEs}

In contrast to the CATEs, QTEs do not identify heterogeneity by subgroups of the population but rather by the potential earnings distributions.\footnote{Under very strong assumptions, the effects on the potential outcome distribution coincide with the individual causal effects. These assumptions imply that individuals do not systematically change their ranks in the potential outcome distributions as a result of treatment status \citep[see the discussion in, e.g.,][]{ch05,fir07}.} The potential outcome distributions are defined by
\begin{equation}\label{eq11}
\begin{array}{rl}
F_{Y(1)}(y) &= Pr(Y_{it}(1) \leq y) = Pr(Y_{it}(0) + \delta_{it} \leq y), \mbox{ and} \\
F_{Y(0)}(y) &= Pr(Y_{it}(0) \leq y) = Pr(Y_{it}(1) - \delta_{it} \leq y),
\end{array}
\end{equation}
with $Y_{it}(1) = Y_{it}(0) + \delta_{it}$. The potential quantile $Q_{Y(d)}(\tau)$ is the minimum value of $Y_{it}(d)$ such that, at minimum, the share $\tau$ of the earnings distribution lies below this value. QTEs are defined as
\begin{equation*}
\delta_{QTE}(\tau)= Q_{Y(1)}(\tau) - Q_{Y(0)}(\tau),
\end{equation*}
the difference between the potential quantiles.

\cite{bit17} propose the simulated potential outcome distributions, 
\begin{equation}
\begin{array}{rl} \label{eq12}
F_{Y(1)}^{S}(y) & = Pr(Y_{it}(0) + \bar{\delta}(X_{it}) \leq y), \mbox{ and} \\
F_{Y(0)}^{S}(y) & = Pr(Y_{it}(1) - \bar{\delta}(X_{it}) \leq y).
\end{array}
\end{equation}
When 
\begin{align}
F_{Y(1)}^{S}(y) & = F_{Y(1)}(y), \label{eq13} \mbox{ and} \\
F_{Y(0)}^{S}(y) & = F_{Y(0)}(y),  \label{eq13b}
\end{align}
then CATEs and QTEs carry the same information and differ only in how they report the effects.\footnote{When CATEs and QTEs are nested, it is always possible to calculate the QTEs from the CATEs using the simulated potential outcome distributions. However, it is not necessarily possible to calculate CATEs from QTEs.} I exploit conditions (\ref{eq13}) and (\ref{eq13b}) to determine whether CATEs and QTEs are nested. Moreover, (\ref{eq13}) and (\ref{eq13b}) are necessary (but not sufficient) conditions for $\delta_{it}=\bar{\delta}(X_{it})$, as can be observed when comparing (\ref{eq11}) and (\ref{eq12}).

\section{Results \label{sec5}}

\subsection{Local constant model}

To create a benchmark, I first document some results for the local constant model. Following one of the main specifications in \cite{bit17}, I stratify the data by previous earnings and quarters elapsed since RA. I classify previous earnings seven quarters before RA into zero earnings ($earn_0$), earnings below the median among those with positive earnings ($earn_1$), and earnings above the median ($earn_2$). Furthermore, I create dummies for each elapsed quarter since RA ($q_1$, ..., $q_7$). Then, I fully interact these dummy variables and the treatment dummy ($D$), such that the model 
\begin{equation*}
Y =  \sum_{k=0}^{2} \sum_{t=1}^{7} \gamma_{kt} \cdot earn_k  \cdot q_t + \sum_{k=0}^{2} \sum_{t=1}^{7} \delta_{kt} \cdot D \cdot earn_k  \cdot q_t ,
\end{equation*}
is fully stratified. The CATEs are $\delta_{kt}$.

Table \ref{res1} reports the percentage of positive and negative CATEs. The first column reports the results for the full sample. 82\% of the CATEs are positive, and 18\% are negative. The second column reports the share of positive and negative CATEs in the sample of AFDC participants. The shares do not change compared to the full sample, which provides reassurance that the predicted CATEs are balanced.

\[ \text{Table~\ref{res1} around here}  \]

Column three of Table \ref{res1} reports the CATE results for unemployed individuals under AFDC. The theoretical labor supply model predicts that I find some positive and no negative CATEs (H1). Indeed, the share of negative CATEs (10\%) is relatively low, and the null hypothesis $H_0^+$ for non-negative CATEs cannot be rejected. Thus, there is empirical support for the theoretical hypothesis H1. 

Column four of Table \ref{res1} reports the CATE results for the AFDC participants with positive earnings below the FPL. For this subsample, the theoretical labor supply model predicts a mix of positive and negative CATEs (H2). However, the null hypothesis $H_0^+$ for non-negative CATEs cannot be rejected. I find empirical evidence only for positive CATEs ($H_0^-$ rejected), which is insufficient to provide evidence for theoretical hypothesis H2. 

Columns five and six of Table \ref{res1} report the CATE results for the AFDC participants with earnings above the FPL. Column five uses the exact number of children to measure the FPL and column six uses the inflated FPL measure considering one ``extra child''. No matter which FPL measure is used, the null hypothesis $H_0^-$ is rejected, suggesting that some individuals with earnings above the FPL have positive CATEs, contradicting theoretical hypothesis H3.

To summarize, the local constant model provides evidence for theoretical hypothesis H1 but rejects H2 and H3.

\subsection{CML results}

\subsubsection{Estimation of the nuisance parameters}

To estimate the nuisance parameters, I always use the random forest estimator with the kitchen sink covariates. In this way, the results of the different CATE estimators do not depend on the specification of the nuisance parameters. 

Figure B.1 in the Online Appendix B documents the histogram of the estimated Jobs First assignment probability. Under RA, I would expect no variation in the assignment probability. The assignment probability varies between 41\% and 61\%, with an average assignment probability of 50\%. Even though the range of the estimated treatment probability is narrow, it is far from homogeneous. However, the modified outcome approach accounts for the differences in the treatment probabilities. Furthermore, I do not have to worry about common support problems because the propensity score is far from zero and one \citep[see discussion in, e.g.,][]{le17}.

Figure B.2 in the Online Appendix B reports the densities of the estimated earnings under the AFDC and Jobs First programs. The estimated earnings are always higher than zero, i.e., they do not capture the mass points at zero earnings (see Figure B.3 in the Online Appendix B).

\subsubsection{Results of the modified outcome approach}

Table \ref{res2} reports the results of the modified outcome approach. A comparison of columns (1) and (2) suggests that restricting the sample to AFDC participants does not alter the results strongly, no matter which estimator is used. %, as is expected under the randomized treatment allocation.

\[ \text{Table~\ref{res2} around here}  \]

The results for trees and forests with baseline controls are qualitatively similar to the findings from the local constant model, suggesting that using a data-driven approach to select subgroups, without incrementing the covariates, does not improve the match between the empirical and theoretical results. I can provide evidence for H1 but must reject H2 and H3.

Using the decent selection of covariates and the tree estimator enables me to detect evidence for the theoretical hypothesis H2 (Table \ref{res2}, column (4)). I find evidence of positive and negative CATEs for the group with positive earnings below the FPL under AFDC. However, I still reject H3. Furthermore, I find evidence for negative CATEs for the group of unemployed AFDC participants. This is not possible because of the limited support for earnings. Using the kitchen sink controls and the tree estimator allows me to provide evidence for H2 and H3, but I still have to reject H1 because of negative CATEs for the unemployed, suggesting that increasing the number of covariates while maintaining the within-group constant effect modeling restriction does not allow me to provide empirical evidence for all theoretical hypotheses.

In Figures E.1-E.3 in the Online Appendix E, I document the relative MSE of the cross-validation samples. For the tree with the baseline controls, the relative MSE is almost flat. Regardless of how the tree estimator stratifies the sample, it never significantly outperforms the benchmark of homogeneous effects. The selected tree minimizes the MSE with 26 final leaves or 25 splits (as opposed to 21 groups in the local constant model). Figure E.4 in the Online Appendix E shows the structure of the tree with baseline controls. Of the 25 splits, 24 are based on previous earnings, and only one is based on the quarters elapsed. The relative MSE of the trees with the decent and kitchen sink covariates is never saturated (see Figures E.2 and E.§ in the Online Appendix E). Eventually, the tree estimators do not generate additional leaves because I impose the restriction that each leaf must have 50 observations, which further indicates that the modeling restriction of the within-group constant effects is not appropriate in the Jobs First case. Figures E.5 and E.6 in the Online Appendix E show the complex structure of the trees with the decent and kitchen sink covariates. Earnings and welfare history, as well as information about children, are important split variables.

The forest and tree results are qualitatively similar when using the decent controls. However, using the forest estimator and the kitchen sink controls enables me to find empirical evidence supporting the theoretical hypotheses H1, H2, and H3. Accordingly, this is the only CML specification that provides evidence for the theoretical labor supply model. For the unemployed under AFDC, the estimated CATEs of the forests with kitchen sink covariates are sometimes positive and otherwise non-negative (Table \ref{res2}, column (3)). This result is empirical evidence supporting the theoretical hypothesis H1. For the group with positive earnings below the FPL under AFDC, the estimated CATEs are sometimes positive and sometimes negative (Table \ref{res2}, column (4)). This result represents empirical evidence for the theoretical hypothesis H2. For the groups with earnings above the FPL under AFDC, the estimated CATEs are sometimes negative and otherwise non-positive (Table \ref{res2}, columns (5) and (6)). This result is empirical evidence for the theoretical hypothesis H3.

\cite{kli16a} report that between 20\% and 100\% of women who do not receive welfare under AFDC reduce their labor supply under Jobs First. The results of the forest estimator with kitchen sink controls suggest 85-90\% negative CATEs for individuals with earnings above the FPL under AFDC. Furthermore, 29\% of the individuals with positive earnings below the FPL have negative CATEs (in this group, not all AFDC recipients are off welfare). This result suggests that the point estimates are roughly in the range of \cite{kli16a}, even without having confidence intervals for labor supply response probabilities.

Figure \ref{fig1x} shows the aggregated CATEs of the forest estimator with kitchen sink controls on the ordinate and the difference between earnings and FPL under AFDC on the abscissa. Figure \ref{fig1x} documents positive effects far below the FPL, negative effects slightly below and above the FPL, and insignificant effects far above the FPL. This finding nicely summarizes the theoretical predictions. Accordingly, CML methods can provide empirical evidence supporting the theoretical predictions of the Jobs First experiment, but I must incorporate many covariates and allow for continuously distributed treatment effects. This finding is consistent with the results of Doriey et al. (2018), who show with their data challenge that as more flexible algorithms model the response surface, their performance is more promising.

\[ \text{Figure~\ref{fig1x} around here}  \]

\subsubsection{Additional results}

Table \ref{res3} reports the results obtained from the causal forest estimator \citep[e.g.,][]{Athey2017a} with the kitchen sink covariates. The results do not differ strongly from the estimates of the modified outcome approach with the random forest and kitchen sink covariates. Both CML methods provide empirical evidence supporting the theoretical labor supply predictions.

\[ \text{Table~\ref{res3} around here}  \]

The early CML literature suggests that the modified outcome method without confounder adjustments
\begin{equation*}
Y_{it}^{**} = \frac{D_i - Pr(D_i=1)}{Pr(D_i=1)Pr(D_i=0)}Y_{it}
\end{equation*}
is sufficient to estimate the CATEs, $\delta(X_{it})=E[Y_i^{**}|X_{it}]$, in randomized experiments. Table \ref{res4} documents that the modified outcome method without covariate adjustments fails to provide evidence for the theoretical labor supply predictions, even for the random forest estimator and kitchen sink controls. This finding is consistent with previous studies documenting the poor properties of the modified outcome method without covariate adjustments \citep[e.g.,][]{ath16}. The shares of positive and negative CATES differ greatly between the first and second columns of Table \ref{res4}, suggesting that the estimated CATEs are not balanced between the Jobs First and AFDC participants. The adjustment for confounders appears crucial (especially when many covariates are incorporated) because even small covariate imbalances could be picked up and misused by the machine learning algorithms.

\[ \text{Table~\ref{res4} around here}  \]

\subsection{Relation between CATEs and QTEs \label{sec6}}

\cite{bit06} use a QTE approach to evaluate the Jobs First program. Figure \ref{fig1c} replicates their main results. QTE can also provide empirical evidence supporting the theoretical labor supply predictions, which raises the following question: do CATEs and QTEs contain the same information?

\[ \text{Figure~\ref{fig1c} around here}  \]

Figure \ref{fig14} reports the simulated earnings distributions using the random forest estimator with the kitchen sink covariates. The simulated earnings distributions are sometimes above and sometimes below the potential earnings distributions. However, the simulated distributions cannot detect the mass point at zero. \cite{heck97} point out, that distributions with differing
mass points cannot be equal. Thus, the hypothesis that the potential and simulated earnings distributions are equal can be formally rejected, which suggests that the QTEs and CATEs contain different information. Furthermore, this finding is evidence that the CATEs are not equal to the individual causal effects.

\[ \text{Figure~\ref{fig14} around here}  \]

In addition, I report Kolmogorov-Smirnov tests for equality of the positive part of the potential and simulated earnings distributions in Online Appendix F. For all estimation approaches, the tests reject that the QTEs and CATEs are nested. 

Whether it is more appropriate to use CATEs or QTEs depends on the concrete research questions. For example, CATEs are more appropriate for developing assignment rules for programs \citep[see, e.g.,][for a discussion]{Athey2017x}. QTEs are more appropriate for investigating earnings inequalities when we are not concerned about the exact locations of specific individuals in the earnings distributions.

\section{Conclusions}

I study the value added by using CML methods in a case study of Connecticut's' Jobs First randomized welfare experiment. In this application, conventional CATE estimators fail to find supporting evidence for the theoretical labor supply predictions. I provide evidence that CML methods can overcome this disadvantage. Accordingly, CML methods can add value to a Jobs First evaluation in the sense that they can provide evidence supporting the theoretical labor supply predictions. However, this strategy works only when the CML methods incorporate many important heterogeneity variables and allow for continuously distributed treatment effects. 

CML methods cannot uncover the entire effect heterogeneity of the Jobs First program. The estimates of the CATEs and QTEs do not contain the same information. Furthermore, CML methods cannot detect mass points in the earnings distributions. Because of the case study style of this research, it is difficult to make statements about the external validity of the results.

\bibliographystyle{econometrica}
\bibliography{bibliothek}

\clearpage

\section*{Figures}

\begin{figure}[h]
	\begin{center}
		\caption{Earnings and welfare transfers under AFDC and Jobs First. \label{fig0}}
		\includegraphics[width=12cm]{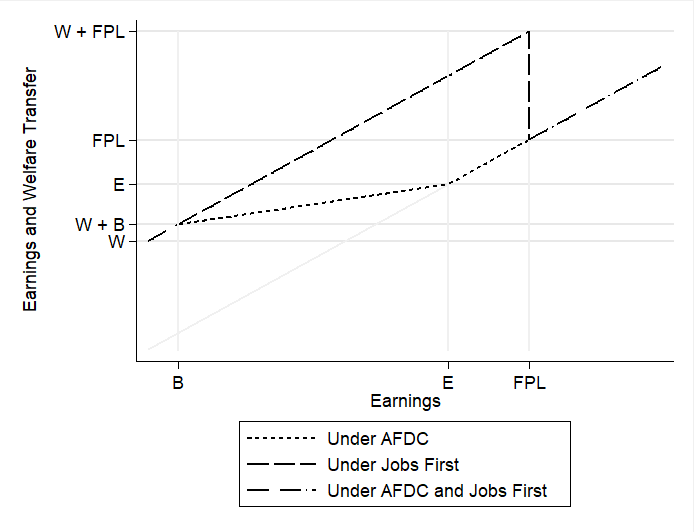}
		\end{center}
		\parbox{\textwidth}{\scriptsize Note: Unemployed persons receive the maximum welfare amount W under AFDC and Jobs First. Under AFDC, all earnings below B are disregarded. Any earnings above B reduce the welfare amount proportionally. Welfare is completely terminated at earnings E. Under Jobs First, all earnings below the FPL are disregarded. Any earnings above the FPL terminate welfare payments.  }
\end{figure}

\begin{figure}[]
	\begin{center}
		\caption{CATEs by difference between earnings and FPL (per quarter). \label{fig1x}}
		\includegraphics[width=12cm]{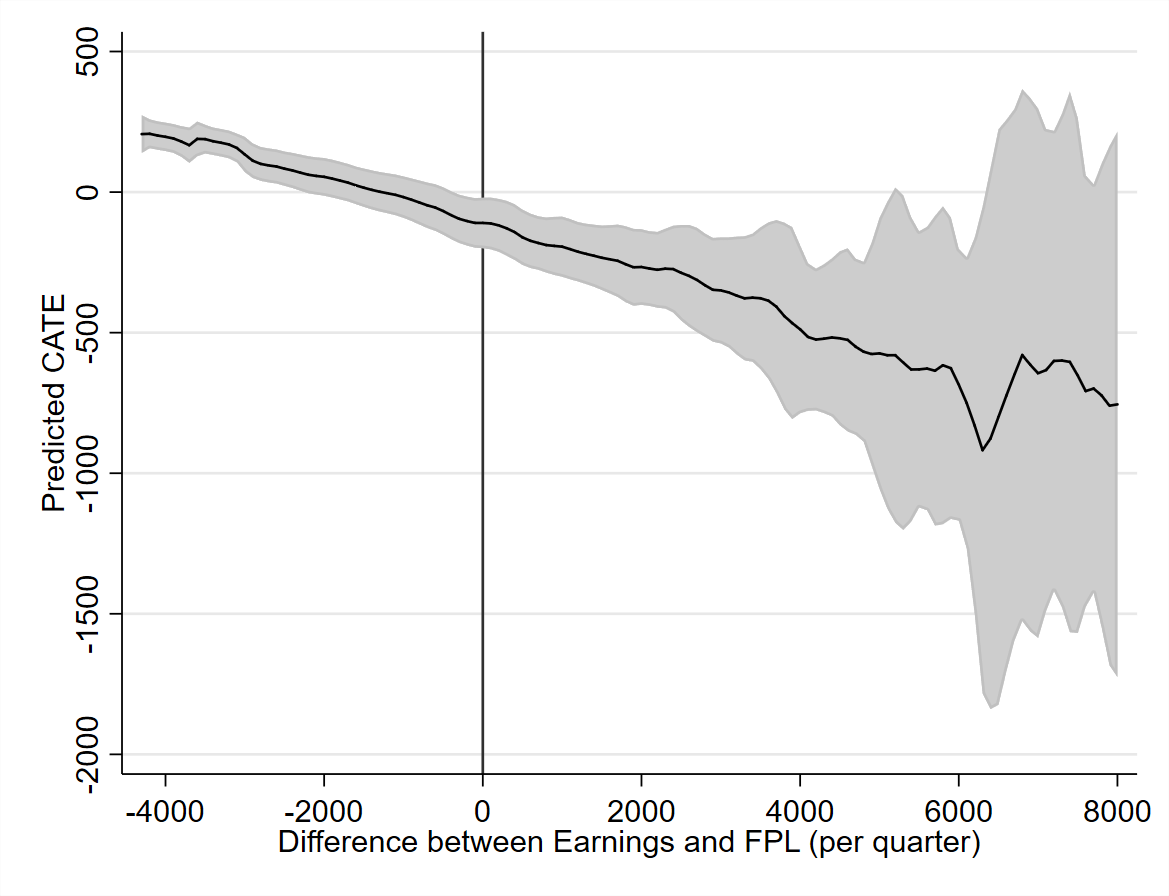}
		\end{center}
		\parbox{\textwidth}{\scriptsize Note: I estimate the aggregated CATEs by the difference between the quarterly earnings and FPL (``extra child'') using a local-constant regression. I use Silverman's rule to specify the bandwidth. FPL (``extra child'') means that I inflate the number of children per mother by one when calculating the assistance unit size. Figure B.4 in Online Appendix B shows the same figure using the FPL without inflating the number of children by one. The gray area reports the 95\% confidence intervals that are estimated using an individual-level clustered bootstrap approach (with 1,999 replications). I control for the kitchen sink covariates. The figure is truncated at 8,000 US dollars.  }
\end{figure}

\begin{figure}[]
	\begin{center}
		\caption{Quantile treatment effects. \label{fig1c}}
		\includegraphics[width=12cm]{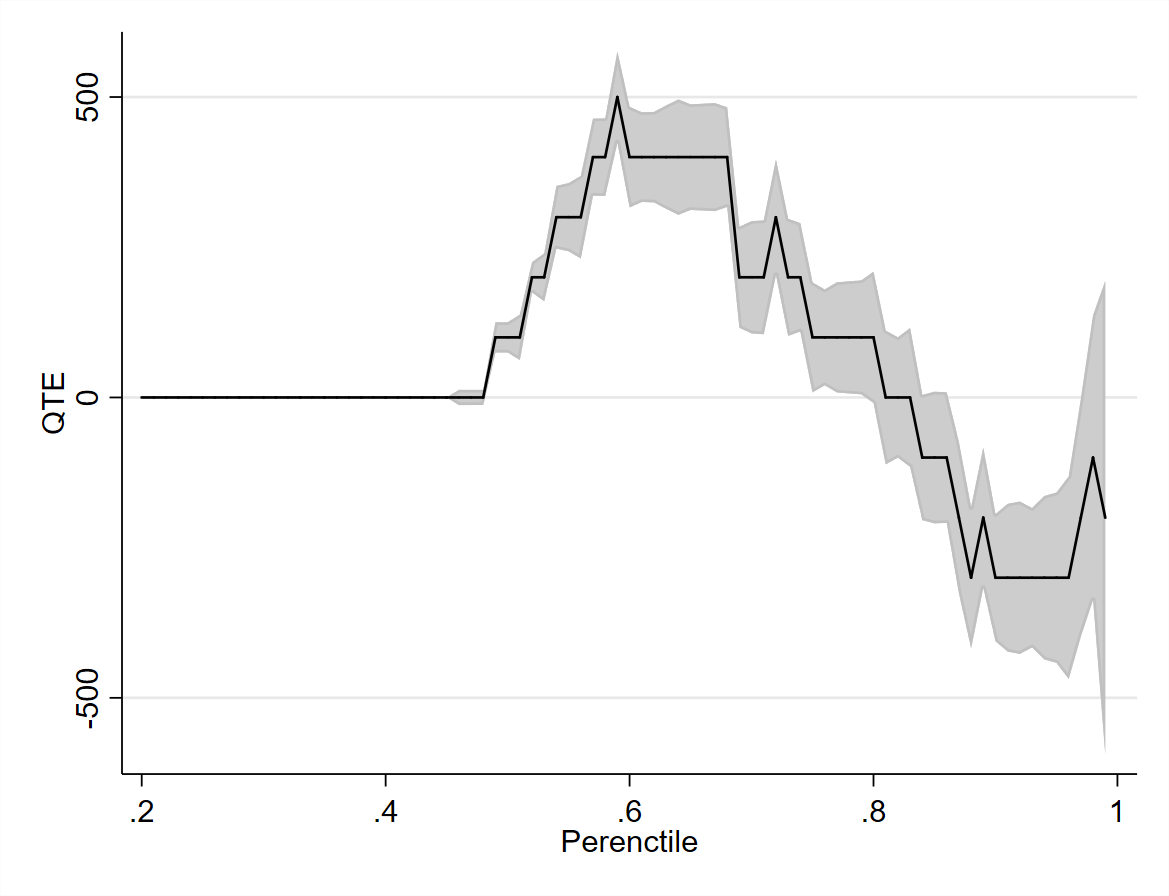}
		\end{center}
		\parbox{\textwidth}{\scriptsize Note: Replication of \cite{bit06}. The gray area shows the 95\% confidence interval.  }
\end{figure}

\begin{figure}[h!]
	\begin{center}
		\caption{Potential and simulated earnings distributions obtained from the random forest estimator with kitchen sink controls. \label{fig14}}
		\begin{tabular}{c}
		\subfigure[Under AFDC]{\includegraphics[width=12cm]{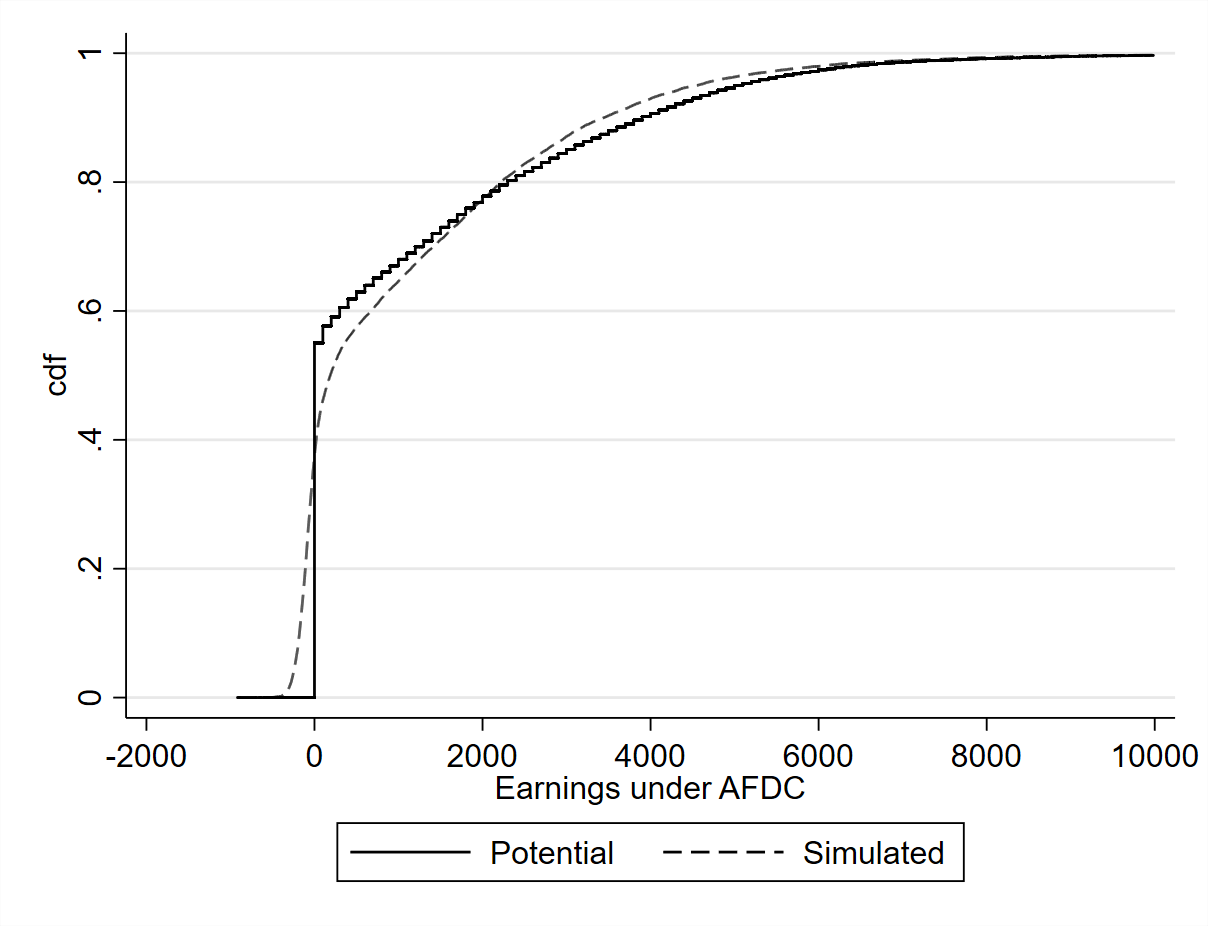}}\\
		\subfigure[Under Jobs First]{\includegraphics[width=12cm]{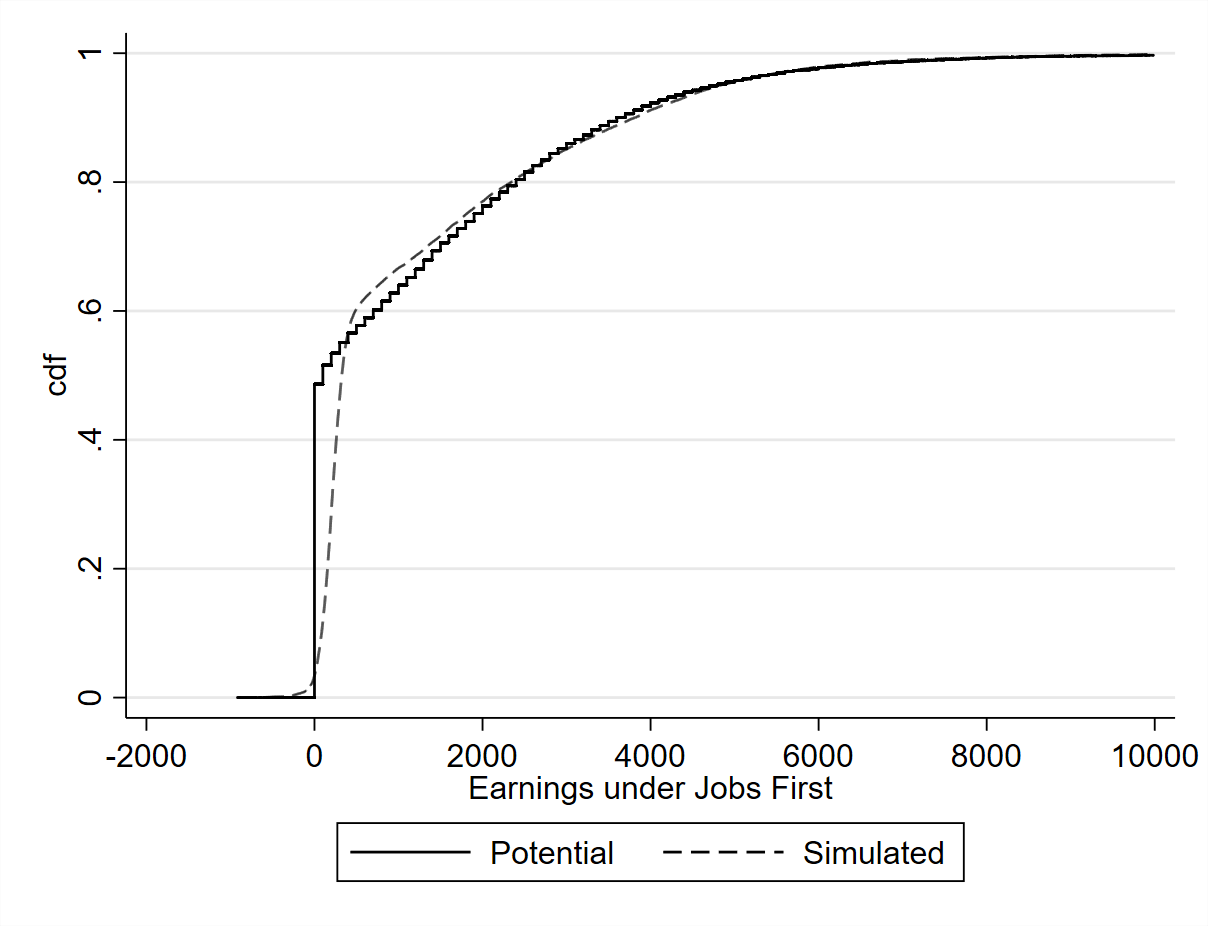}}
		\end{tabular}
		\end{center}
	\end{figure}

\clearpage

\section*{Tables}

\begin{table}[h!]
\begin{center} 
\caption{List of covariates.\label{list}}
\begin{tabularx}{\textwidth}{lX}\hline \hline
Baseline: & quarters elapsed since RA, earnings seven quarters prior to RA\\\hline
Decent: & age, education, number of children, age of the youngest child, amount of AFDC assistance received seven quarters prior to RA, amount of food stamps received seven quarters prior to RA, dummy variable indicating a positive amount of earnings in at least one of the seven quarters prior to RA, dummy variable indicating a positive amount of AFDC assistance received in at least one of the seven quarters prior to RA, dummy variable indicating a positive amount of food stamps in at least one of the seven quarters prior to RA \\\hline
Kitchen sink: & ethnicity, marital status, city of residence, information on residence in a publicly subsidized home, information on relocations, participation in different types of education and labor market programs in the 12 months prior to RA (e.g., English as a Secondary Language (ESL), Adult Basic Education (ABE), General Education Development (GED), job readiness skills, work experience, vocational education, post-secondary education, and high school), earnings for each of the seven quarters prior to the RA, the amount of AFDC assistance received for each of the seven quarters prior to the RA, the amount of food stamps received for each of the seven quarters prior to RA, the number of quarters on AFDC, a dummy variable indicating whether the family received AFDC during childhood, a dummy variable indicating whether work was never recorded, a dummy variable indicating whether work was recorded at RA \\\hline \hline
\end{tabularx}
\parbox{\textwidth}{\scriptsize Note: The decent covariates also include the baseline covariates. The kitchen sink covariates also include the decent covariates. I include dummies for missing values whenever necessary (see Table B.2 in the Online Appendix B for details). }
\end{center}
\end{table}

\begin{table}[h!]
  \begin{center}
  \caption{Descriptive statistics of the main variables. \label{desc}}
    \begin{tabularx}{\textwidth}{Xccccc}\hline\hline
          & \multicolumn{2}{c}{Jobs First} & \multicolumn{2}{c}{AFDC} & SD \\
         & Mean  & St. Dev. & Mean  & St. Dev. &  \\
    \cline{2-6}       & (1)   & (2)   & (3)   & (4)   & (5) \\\hline
    Earnings per quarter (in \textdollar) & 1173  & 1789  & 1125  & 1868  & 2.6 \\
		Share of participants with &       &       &       &       &  \\
   \qquad no earnings & 0.49  & 0.50  & 0.55  & 0.50  & 13.3 \\
   \qquad earn. below FPL & 0.39  & 0.49  & 0.31  & 0.46  & 17.0 \\
   \qquad earn. above FPL & 0.13  & 0.33  & 0.14  & 0.35  & 4.1 \\
   \qquad earn. above FPL (``extra child'') & 0.09  & 0.28  & 0.10  & 0.30  & 4.4 \\\hline
    \multicolumn{6}{c}{Baseline covariates} \\\hline
    Quarters since RA & 4.0   & 2.0   & 4.0   & 2.0   & 0.0 \\
    Earnings in pre-Q7 (in \textdollar) & 682   & 1552  & 774   & 1781  & 5.5 \\\hline
    \multicolumn{6}{c}{Decent covariates} \\\hline
    Age categories &       &       &       &       &  \\
  \qquad $<$ 20 years & 0.09  & 0.28  & 0.09  & 0.28  & 1.2 \\
   \qquad 20-24 years & 0.20  & 0.40  & 0.21  & 0.41  & 2.7 \\
  \qquad 25-34 years & 0.41  & 0.49  & 0.42  & 0.49  & 1.5 \\
   \qquad 35-45 years & 0.25  & 0.43  & 0.23  & 0.42  & 4.1 \\
   \qquad $>$ 44 years & 0.05  & 0.22  & 0.06  & 0.23  & 1.2 \\
    Education categories &       &       &       &       &  \\
   \qquad No degree & 0.33  & 0.47  & 0.31  & 0.46  & 3.8 \\
   \qquad High school & 0.55  & 0.50  & 0.57  & 0.50  & 3.2 \\
   \qquad More than high school & 0.06  & 0.24  & 0.06  & 0.23  & 1.9 \\
    Age of youngest child (in years) & 4.6   & 4.7   & 4.5   & 4.8   & 2.2 \\
    Number of children & 1.6   & 1.0   & 1.5   & 1.0   & 6.0 \\
    AFDC pre-Q7 (in \textdollar) & 920   & 925   & 865   & 896   & 6.0 \\
    Food stamps pre-Q7 (in \textdollar) & 306   & 319   & 293   & 301   & 4.4 \\
    Any earnings pre-Q1/7 & 0.33  & 0.37  & 0.36  & 0.38  & 7.9 \\
    Any AFDC pre-Q1/7 & 0.57  & 0.45  & 0.54  & 0.45  & 6.5 \\
    Any food stamps pre-Q1/7 & 0.61  & 0.44  & 0.60  & 0.43  & 2.1 \\\hline
		  Participants   & \multicolumn{2}{c}{2,396} & \multicolumn{2}{c}{2,406} &\\
    Observations   & \multicolumn{2}{c}{16,772} & \multicolumn{2}{c}{16,842} &  \\\hline\hline
    \end{tabularx}
		\end{center}
		\parbox{\textwidth}{\scriptsize Note: The last column reports the standardized difference (SD).  Earnings in pre-Q7 refers to earnings in the seven quarters before RA. Any earnings pre-Q1/7 is a dummy variable indicating that earnings were positive in at least one of the seven quarters prior to RA. FPL (``extra child'') means that I inflate the number of children per mother by one when calculating the assistance unit size.}
		\end{table}

\begin{table}[]
  \begin{center}
 \caption{Results of the local constant model. \label{res1}}
    \begin{tabular}{lcccccc}
   \hline\hline
          &   & \multicolumn{5}{c}{Under AFDC} \\
      \cline{3-7}    & Full &    & & Pos. earn. & Earn. & Earn. above \\
          &   Sample    &  All     & Unempl.       & below  & above  & FPL (``extra  \\
					          &       &       &       & FPL & FPL &  child'') \\
        \cline{2-7}  & (1)   & (2)   & (3)   & (4)   & (5)   & (6) \\
    \hline
    \multicolumn{7}{c}{Local constant model with baseline covariates} \\
   \hline
    Positive CATEs & 82\%  & 82\%  & 90\%  & 79\%  & 56\%  & 53\% \\
    Negative CATEs & 18\%  & 18\%  & 10\%  & 21\%  & 44\%  & 47\% \\
    p-value $H_0^+$ & 0.29  & 0.29  & 0.56  & 0.12  & 0.00  & 0.00 \\
    p-value $H_0^-$ & 0.00  & 0.00  & 0.00  & 0.00  & 0.00  & 0.00 \\
    \hline
    Observations & 33,621 & 16,842 & 8,988  & 4,967  & 2,313  & 1,713 \\
    \hline
    \hline
    \end{tabular}
\end{center}
		\parbox{\textwidth}{\scriptsize Note: $H_0^+$ is the null hypothesis that all CATEs are non-negative. $H_0^-$ is the null hypothesis that all CATEs are non-positive. FPL (``extra child'') means that I inflate the number of children per mother by one when calculating the assistance unit size. P-values are calculated with an individual-level clustered bootstrap procedure (with 1,999 replications). The details of the distributional tests are provided in the Online Appendix D. }
\end{table}

\begin{table}[h!]
  \begin{center}
  \caption{Results of the modified covariate approach. \label{res2}}
    \begin{tabular}{lcccccc}
   \hline\hline
          &   & \multicolumn{5}{c}{Under AFDC} \\
      \cline{3-7}    & Full &    & & Pos. earn. & Earn. & Earn. above \\
          &   Sample    &  All     & Unempl.       & below  & above  & FPL (``extra  \\
					          &       &       &       & FPL & FPL &  child'') \\
        \cline{2-7}  & (1)   & (2)   & (3)   & (4)   & (5)   & (6) \\
    \hline
    \multicolumn{7}{c}{Tree with baseline covariates} \\
   \hline
    Positive CATEs & 88\%  & 87\%  & 93\%  & 85\%  & 68\%  & 66\% \\
    Negative CATEs & 12\%  & 13\%  & 7\%   & 15\%  & 32\%  & 34\% \\
    p-value $H_0^+$ & 0.32  & 0.29  & 0.45  & 0.27  & 0.06  & 0.04 \\
    p-value $H_0^-$ & 0.00  & 0.00  & 0.00  & 0.00  & 0.01  & 0.01 \\
    \hline
    \multicolumn{7}{c}{Tree with decent covariates} \\
    \hline
    Positive CATEs & 53\%  & 53\%  & 60\%  & 50\%  & 35\%  & 33\% \\
    Negative CATEs & 47\%  & 47\%  & 40\%  & 50\%  & 65\%  & 67\% \\
    p-value $H_0^+$ & 0.00  & 0.00  & 0.00  & 0.00  & 0.00  & 0.00 \\
    p-value $H_0^-$ & 0.00  & 0.00  & 0.00  & 0.00  & 0.01  & 0.07 \\
    \hline
    \multicolumn{7}{c}{Tree with kitchen sink covariates} \\
    \hline
    Positive CATEs & 55\%  & 56\%  & 67\%  & 51\%  & 29\%  & 30\% \\
    Negative CATEs & 45\%  & 44\%  & 33\%  & 49\%  & 71\%  & 70\% \\
    p-value $H_0^+$ & 0.00  & 0.00  & 0.00  & 0.00  & 0.00  & 0.00 \\
    p-value $H_0^-$ & 0.00  & 0.00  & 0.00  & 0.00  & 1.00  & 1.00 \\
    \hline
    \multicolumn{7}{c}{Forest with baseline covariates} \\
    \hline
    Positive CATEs & 84\%  & 83\%  & 91\%  & 79\%  & 58\%  & 55\% \\
    Negative CATEs & 16\%  & 17\%  & 9\%   & 21\%  & 42\%  & 45\% \\
    p-value $H_0^+$ & 0.52  & 0.47  & 0.77  & 0.32  & 0.00  & 0.00 \\
    p-value $H_0^-$ & 0.00  & 0.00  & 0.00  & 0.00  & 0.00  & 0.00 \\
   \hline
    \multicolumn{7}{c}{Forest with decent covariates} \\
   \hline
    Positive CATEs & 69\%  & 70\%  & 86\%  & 63\%  & 20\%  & 15\% \\
    Negative CATEs & 31\%  & 30\%  & 14\%  & 37\%  & 80\%  & 85\% \\
    p-value $H_0^+$ & 0.00  & 0.00  & 0.00  & 0.00  & 0.00  & 0.00 \\
    p-value $H_0^-$ & 0.00  & 0.00  & 0.00  & 0.00  & 1.00  & 1.00 \\
    \hline
    \multicolumn{7}{c}{Forest with kitchen sink covariates} \\
    \hline
    Positive CATEs & 75\%  & 75\%  & 94\%  & 71\%  & 15\%  & 10\% \\
    Negative CATEs & 25\%  & 25\%  & 6\%   & 29\%  & 85\%  & 90\% \\
    p-value $H_0^+$ & 0.00  & 0.00  & 0.12  & 0.00  & 0.00  & 0.00 \\
    p-value $H_0^-$ & 0.00  & 0.00  & 0.00  & 0.00  & 1.00  & 1.00 \\
   \hline
    Observations & 33,621 & 16,842 & 8,988  & 4,967  & 2,313  & 1,713 \\
    \hline
    \hline
    \end{tabular}
\end{center}
		\parbox{\textwidth}{\scriptsize Note: $H_0^+$ is the null hypothesis that all CATEs are non-negative. $H_0^-$ is the null hypothesis that all CATEs are non-positive. FPL (``extra child'') means that I inflate the number of children per mother by one when calculating the assistance unit size. P-values are calculated with an individual-level clustered bootstrap procedure (with 1,999 replications). The details of the distributional tests are provided in the Online Appendix D. }
\end{table}

\begin{table}[]
  \begin{center}
  \caption{Results of the causal forest approach. \label{res3}}
    \begin{tabular}{lcccccc}
   \hline\hline
          &   & \multicolumn{5}{c}{Under AFDC} \\
      \cline{3-7}    & Full &    & & Pos. earn. & Earn. & Earn. above \\
          &   Sample    &  All     & Unempl.       & below  & above  & FPL (``extra  \\
					          &       &       &       & FPL & FPL &  child'') \\
        \cline{2-7}  & (1)   & (2)   & (3)   & (4)   & (5)   & (6) \\
    \hline
    \multicolumn{7}{c}{Causal forest with kitchen sink covariates} \\
   \hline
    Positive CATEs & 78\%  & 77\%  & 93\%  & 74\%  & 25\%  & 20\% \\
    Negative CATEs & 22\%  & 23\%  & 7\%   & 26\%  & 75\%  & 80\% \\
    p-value $H_0^+$ & 0.00  & 0.00  & 0.34  & 0.00  & 0.00  & 0.00 \\
    p-value $H_0^-$ & 0.00  & 0.00  & 0.00  & 0.00  & 1.00  & 1.00 \\
		    \hline
    Observations & 33,621 & 16,842 & 8,988  & 4,967  & 2,313  & 1,713 \\
    \hline
    \hline
    \end{tabular}
\end{center}
		\parbox{\textwidth}{\scriptsize Note: $H_0^+$ is the null hypothesis that all CATEs are non-negative. $H_0^-$ is the null hypothesis that all CATEs are non-positive. FPL (``extra child'') means that I inflate the number of children per mother by one when calculating the assistance unit size. P-values are calculated with an individual-level clustered bootstrap procedure (with 1,999 replications). The details of the distributional tests are provided in the Online Appendix D. }
\end{table}

\begin{table}[h]
  \begin{center}
  \caption{Results of the modified outcome approach without confounder adjustment. \label{res4}}
    \begin{tabular}{lcccccc}
   \hline\hline
          &   & \multicolumn{5}{c}{Under AFDC} \\
      \cline{3-7}    & Full &    & & Pos. earn. & Earn. & Earn. above \\
          &   Sample    &  All     & Unempl.       & below  & above  & FPL (``extra  \\
					          &       &       &       & FPL & FPL &  child'') \\
        \cline{2-7}  & (1)   & (2)   & (3)   & (4)   & (5)   & (6) \\
    \hline
    \multicolumn{7}{c}{Forest with kitchen sink covariates} \\
   \hline
    Positive CATEs & 68\%  & 50\%  & 74\%  & 29\%  & 3\%   & 2\% \\
    Negative CATEs & 32\%  & 50\%  & 26\%  & 71\%  & 97\%  & 98\% \\
    p-value $H_0^+$ & 0.00  & 0.00  & 0.00  & 0.00  & 0.00  & 0.00 \\
    p-value $H_0^-$ & 0.00  & 0.00  & 0.00  & 0.00  & 1.00  & 1.00 \\\hline
    Observations & 33,621 & 16,842 & 8,988  & 4,967  & 2,313  & 1,713 \\
    \hline
    \hline
    \end{tabular}
\end{center}
		\parbox{\textwidth}{\scriptsize Note: $H_0^+$ is the null hypothesis that all CATEs are non-negative. $H_0^-$ is the null hypothesis that all CATEs are non-positive. FPL (``extra child'') means that I inflate the number of children per mother by one when calculating the assistance unit size. P-values are calculated with an individual-level clustered bootstrap procedure (with 1,999 replications). The details of the distributional tests are provided in the Online Appendix D. }
\end{table}

\end{document}